\documentclass[12pt]{JHEP3}
\usepackage{graphicx}
\usepackage{amsmath}

\newcommand{\rf}[1]{(\ref{#1})}
\newcommand{\beq}{\begin{equation}}
\newcommand{\eeq}{\end{equation}}
\newcommand{\bea}{\begin{eqnarray}}
\newcommand{\eea}{\end{eqnarray}}

\newcommand{\e}{\mbox{e}}
\renewcommand{\d}{\mbox{d}}

\renewcommand{\l}{\lambda}

\renewcommand{\a}{\alpha}

\newcommand{\del}{\delta}

\newcommand{\dg}{\dagger}

\newcommand{\ra}{\rangle}
\newcommand{\la}{\langle}
\newcommand{\prt}{\partial}
\newcommand{\mi}{\!-\!}

\newcommand{\pl}{\!+\!}

\newcommand{\cD}{{\cal D}}

\newcommand{\cG}{{\cal G}}

\newcommand{\tG}{{\tilde{G}}}
\newcommand{\tg}{{\tilde{g}}}

\newcommand{\trho}{{\tilde{\rho}}}
\newcommand{\tl}{{\tilde{l}}}
\newcommand{\tx}{{\tilde{x}}}

\newcommand{\tti}{{\tilde{t}}}
\newcommand{\tH}{{\tilde{H}}}
\newcommand{\tPsi}{{\tilde{\Psi}}}

\newcommand{\hH}{{\hat{H}}}

\newcommand{\bx}{{\bar{x}}}

\newcommand{\sla}{\sqrt{\l}}

\newcommand{\hWg}{{{\hat{W}}_{\l,g}}}

\newcommand{\vac}{|0\ra}
\newcommand{\cav}{\la 0 |}
\newcommand{\dll}{\frac{dl}{l}}

\title{A String Field Theory based on Causal Dynamical 
Triangulations}

\author{Jan Ambj\o rn$^{a,b}$ and Renate Loll$^b$\\ 
$^a$The Niels Bohr Institute, Copenhagen University,
Blegdamsvej 17, DK-2100 Copenhagen \O , Denmark,\\
$^b$Institute for Theoretical Physics, Utrecht University, 
Leuvenlaan 4, NL-3584 CE Utrecht, The Netherlands\\
Email: \email{ambjorn@nbi.dk},~~\email{loll@phys.uu.nl}}
\author{Yoshiyuki Watabiki\\ Tokyo Institute of Technology,
Dept. of Physics, High Energy Theory Group,
2-12-1 Oh-okayama, Meguro-ku,
Tokyo 152-8551, Japan\\
Email: \email{watabiki@th.phys.titech.ac.jp}} 
\author{Willem Westra\\Department of Physics, University of Iceland, 
Dunhaga 3, 107 Reykjavik, Iceland\\
Email: \email{willem@raunvis.hi.is}}
\author{Stefan Zohren\\Blackett Laboratory, Imperial College,
London SW7 2AZ, UK, and\\
Department of Physics, Ochanomizu university, Otsuka, Bunkyo-ku, 
Tokyo 112-8610, Japan\\
Email: \email{stefan.zohren@imperial.ac.uk}}

\abstract{We formulate the string field theory in zero-dimensional target
space corresponding to   
the two-dimensional quantum gravity theory defined through
Causal Dynamical Triangulations. This third quantization
of the quantum gravity theory allows
us in principle to calculate the transition amplitudes
of processes in which the topology of space changes in time, 
and to include non-trivial topologies of space-time.
We formulate the corresponding Dyson-Schwinger equations
and illustrate how they can be solved iteratively.}


\begin{document}

\section{Introduction}\label{intro}

The formulation of a non-critical string theory %%
in which the conformal mode plays an important role dates
back to Polyakov. He emphasized the worldsheet %%
formulation of string theory as a two-dimensional quantum gravity theory
coupled to matter \cite{polyakov}. This triggered non-perturbative
definitions of non-critical string theory 
\cite{ambjorn,david,mkk}, introducing what is now called
dynamical triangulation (DT) as a regularization of the worldsheet
theory. When the dimension of space-time was larger than 1
these attempts in some sense
did not work. One could show that the outcome was not 
a proper string theory, but a theory  where the 
worldsheet had degenerated into branched polymers \cite{ad}.
However, when considering matter fields with central charge
$c \leq 1$, these regularized theories led to what is now
known as non-critical string theory, a very
useful toy model of real string theory. In particular,
it has been possible to formulate a string field theory
of non-critical string theory \cite{sft,moresft} which is 
very much simpler than the critical string field theory.

The use of causal dynamical triangulations (CDT) rather than DT as
a regularization of quantum gravity was inspired by earlier ideas in
\cite{tei}: one insists, starting from a Lorentzian space-time,
that only causal histories contribute to the quantum gravitational 
path integral. In addition, one assumes the presence of a 
global time-foliation.
In this way the space-times appearing in the {\it regularized}
path integral become
a set of piecewise linear causal geometries, made out of triangles
(two-simplices) whose
edge lengths provide an ultraviolet cut-off. For a detailed
description of how to construct these geometries we refer to \cite{al,ajl2d}
in two dimensions and \cite{blp} in higher dimensions.
In order to perform the summation over these causal geometries 
we perform a rotation to Euclidean space-times. Each piecewise linear
causal geometry as defined in \cite{blp} has a continuation
to Euclidean signature, but the class of Euclidean geometries included
in the path integral will only be a subclass of the total class 
of Euclidean geometries, and the result of the summation will therefore be different
from that of Euclidean quantum gravity.

One is interested in the limit where the lattice spacing $a$ goes to zero.
There is evidence for the existence of an underlying 
(non-perturbatively defined) continuum quantum field theory in four 
dimensions \cite{4d} and the results seem to be in qualitative agreement
with recent renormalization group calculations \cite{rg}. 
These intriguing developments in the {\it four}-dimensional theory
are based on numerical simulations, since analytical tools are
presently unavailable. In two dimensions the situation is different,
since the quantum
gravity model can be solved analytically at the discretized level and
the limit $a \to 0$ can be constructed.

In \cite{alwz} we showed that the original two-dimensional CDT model
of quantum gravity defined and solved in \cite{al} can be generalized to a model %%
where one allows for the creation of so-called baby universes, branching
off from the ``parent universe". The creation
of a baby universe results in at least one point where 
from a Lorentzian point of view the metric is degenerate \cite{louko}.
One cannot invoke the classical theory to decide a priori whether or not
such geometries should be included in the path integral.  In
\cite{al} we made the choice to suppress these configurations.
We could also show that if they were completely unsuppressed
one would recover Euclidean 2d quantum gravity as defined via DT
or quantum Liouville theory. The converse was demonstrated in \cite{ackl}:
if one integrates out all baby universes in Euclidean quantum 
gravity, one obtains CDT. 

Quite surprisingly, there exists yet
a third possibility, namely, a double-scaling limit where the creation 
of baby universes in CDT can be associated with the gravitational 
coupling constant \cite{alwz}. In this double-scaling limit 
one can calculate %%
the disc amplitude and finds a result which is analytically
connected to the old CDT result, the expansion parameter
being the gravitational coupling constant. However, this cannot -- at least not
by simple analytic continuation -- be connected to the Euclidean
theory. Thus we have arrived at a theory which allows the 
creation of baby universes, but in a much more controlled way 
than in Euclidean quantum gravity. Of course, unlike the original CDT
prescription, this construction contains causality-violating features at the level of
the piecewise linear Lorentzian geometries. However, as we will see, 
the Lorentzian structure still plays a role in ``taming" them. -- Apart from the interesting
observation that such a new theory exists, it may
have important implications for the higher-dimensional theories.
The attempt to formulate {\it Euclidean} higher-dimensional quantum gravity 
theories using DT as a regularization ran into the problem that baby
universes completely dominate the path integral and make
it difficult to obtain a physically sensible continuum limit. 
Now we see that there may exist a way to include the 
creation of baby universes in a controlled manner, starting with the {\it C}DT 
regularization of the quantum gravity theory.

In this paper we show that the construction of \cite{alwz} %%
can be turned into a full-fledged third quantization  %%
of 2d quantum gravity. In the terminology of \cite{sft} this is 
a string field theory for $c=0$, in the sense that 
it allows the calculation of amplitudes for splitting and
joining of (spatial) universes and as well as the inclusion of
different space-time topologies.

The remainder of this article is organized as follows: In Sec.\ \ref{cap}
we review briefly the results of the generalized CDT model.
In Sec.\ \ref{sft} we show how to define a string field theory, and in 
Sec.\ \ref{alpha} we show how it reproduces the results
of the generalized CDT model. 
In Sec.\ \ref{ds} we
derive the general Dyson-Schwinger equations and 
in Sec.\ \ref{application} we show
how they can be used to calculate in a systematic way 
multi-universe and topology-changing amplitudes. 
Finally, we discuss the interpretation and
possible generalizations in Sec.\ \ref{discuss}.

\section{Generalized Causal Dynamical Triangulation in 2d}\label{cap}

We will initially assume 
that the two-dimensional space-time has topology $S^1\times [0,1]$.
After rotation to Euclidean signature, the pure gravity action is given by
\beq\label{2.a}
S[g_{\mu\nu}] = \l \int \d^2 \xi  \sqrt{\det g_{\mu\nu}(\xi)} +
x \oint \d l_1 + y\oint \d l_2,
\eeq
where $\l$ is  the cosmological constant, 
$x$ and $y$ are two so-called boundary
cosmological constants, $g_{\mu\nu}$ is the metric of a geometry
of the kind described above, and the line integrals refer to
the lengths of the in- and out-boundaries induced by $g_{\mu\nu}$.
The so-called {\it proper-time propagator} is defined by
\beq\label{2.a0}
G_\l (x,y;t) =
\int \cD [g_{\mu\nu}] \; e^{-S[g_{\mu\nu}]}.
\eeq
This represents the Euclideanization of a functional integral over 
space-times with {\it Lorentzian} signature, performed 
over all causal geometries $[g_{\mu\nu}]$
such that the final (or ``exit'') boundary with boundary cosmological constant
$y$ is separated\footnote{The statement that the exit boundary is separated by a 
geodesic distance $t$ from the entrance boundary means
in this context that {\it all} points on the exit 
boundary have a geodesic distance $t$ to the entrance boundary. The 
geodesic distance of a point on the exit loop to the entrance loop
is defined as the 
minimal geodesic distance from the exit point to points on 
the entrance loop. In the piecewise flat, triangulated geometries we are working with,
``distance" is given by ``link distance". In the case of the original, ``pure" 
CDT without any causality violations, the notion of distance between boundaries
just introduced is {\it symmetric} under exchange of entrance and exit boundary.
For the generalized models discussed below, this will no longer be the case.\label{note1}} 
a geodesic distance $t$ from the initial 
(or ``entrance'') boundary
with boundary cosmological constant $x$. To arrive at the integral \rf{2.a0},
all causal geometries have been rotated to Euclidean signature, 
a procedure which is well defined in the CDT regularization of the
path integral.
 
Calculating the path integral \rf{2.a0} with the help of the 
CDT regularization and taking
the continuum limit as the edge length $a$ of the triangles goes
to zero leads to the equation
\beq\label{2.n1}
\frac{\prt}{\prt t} G_\l(x,y;t) = - 
\frac{\prt}{\prt x} \Big[(x^2-\l) G_\l(x,y;t)\Big],
\eeq
which can readily be solved \cite{al}.
Let $l_1$ denote the length of the initial and $l_2$ the
length of the final boundary. Rather than considering a situation
where the boundary cosmological constant $x$ is fixed, we will
take $l_1$ as fixed, and denote the corresponding propagator by
$G_\l (l_1,y;t)$, with similar definitions for $G_\l(x,l_2;t)$
and $G_\l(l_1,l_2;t)$. All of them are related by Laplace transformations,
for instance,
\beq\label{2.a5}
G_\l(x,y;t)= \int_0^\infty \d l_2 \int_0^\infty \d l_1\;
G_\l(l_1,l_2;t) \;\e^{-xl_1-yl_2}.
\eeq

Next, we will turn our attention to the so-called {\it disc amplitude},
associated with a piece of space-time which has the topology of
a disc. Strictly speaking, the disc amplitude does not exist in CDT. 
A spatial slice in a two-dimensional Lorentzian 
space-time of the type we are considering will 
by construction be a one-dimensional space-like subspace of
topology $S^1$, i.e. a circle. Now,
there is no way this can be extended to a well-defined
Lorentzian geometry everywhere in the interior of any finite disc whose boundary is
the circle. The light cones of the geometry must degenerate in at
least a point, because the disc does not extend infinitely in time. 
However, after rotation to Euclidean signature\footnote{We should emphasize
that there is in principle a choice involved when generalizing the unique
Wick rotation of
CDT \cite{al,ajl2d,blp} to situations where the causal structure of the
piecewise flat geometries has singularities. One might 
attach certain complex and/or singular weights to such singularities
in the Euclideanization, for example, of the kind envisaged in \cite{louko}.
When the disc amplitude was first introduced in two-dimensional CDT in
order to compare it to Euclidean models \cite{al}, no extra weight was
associated with it, leading to the disc amplitude \rf{2x.1}. In the present work, 
following \cite{lw,alwz}, we will associate
finite, real weights with baby universes and branching points, as will be
explained in detail below.}, we can {\it define} a disc
amplitude, which is related to $G_\l (x,l_2;t)$ by
\beq\label{2x.1}
W_\l(x) = \int_0^\infty dt \; G_\l (x, l_2=0;t) = \frac{1}{x+\sla}.
\eeq
There is clearly a latest time $t$ where the spatial universe 
contracts to length zero and vanishes into the ``vacuum''.
When introducing the string field theory below, we will see that this process has 
a natural realization as a tadpole term in the string field 
Hamiltonian. 

\FIGURE[t]{
\centerline{\scalebox{0.45}{\rotatebox{0}{\includegraphics{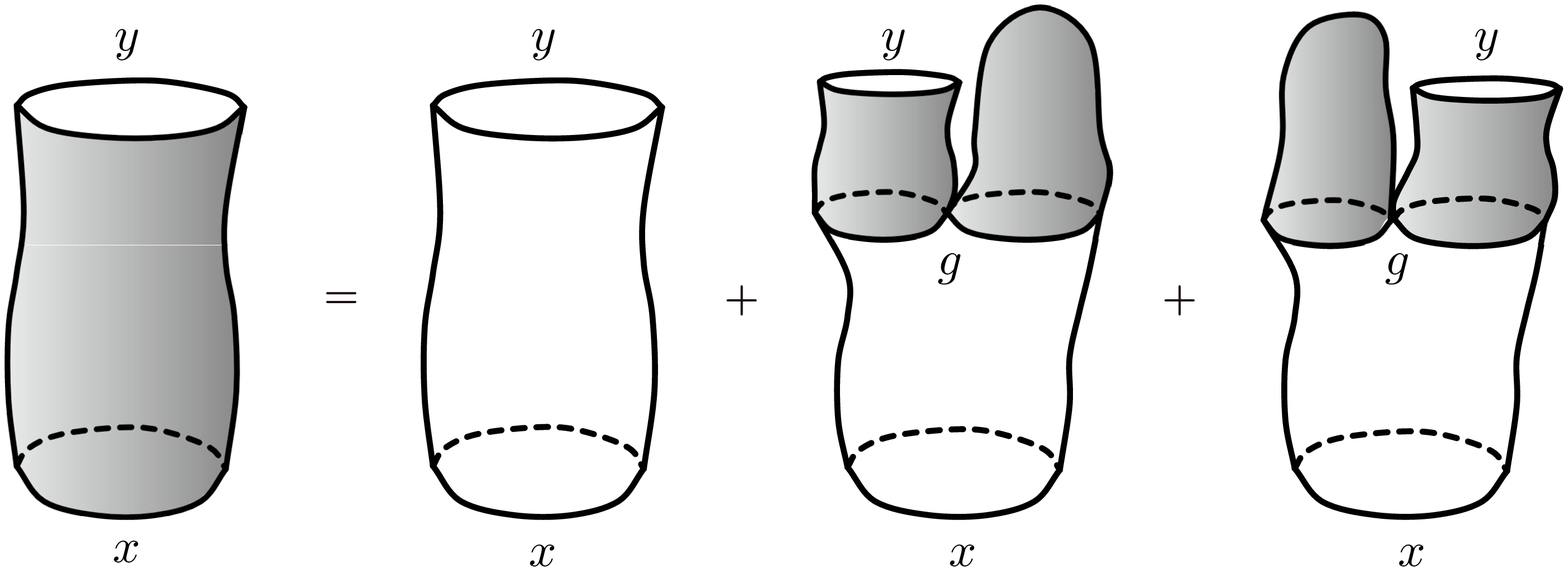}}}}
\caption[fig2]{{\small 
In all four graphs, the geodesic distance from the final to the initial 
loop is given by $t$. Differentiating
with respect to $t$ leads to eq.\ \rf{2.55}. Shaded parts of graphs represent
the full, $g$-dependent propagator $G_{\l,g}$ and disc amplitude $W_{\l,g}$, 
and non-shaded parts the CDT propagator $G_\l$.}}
\label{fig2}
}
We will now allow for the possibility that space 
branches into disconnected parts as a function of proper time $t$,
and introduce a coupling constant $g$ of mass dimension 3 
associated with the branching\footnote{One could in principle have 
considered a more general branching process, 
where more than one baby universe can sprout at 
any given time step $t$. However, starting 
with the discretized theory and a lattice cut off, one can show that such
processes are suppressed when the lattice spacing goes to 
zero \cite{alwz}. This is related to the fact that $g$ 
has mass dimension 3.}. As shown in \cite{alwz},
this modifies the equation for the proper-time propagator to  
\beq\label{2.55}
 \frac{\prt}{\prt t} G_{\l,g}(x,y;t) = 
- \frac{\prt}{\prt x} \Big[\Big((x^2-\l)+2 g\;W_{\l,g}(x)\Big) 
G_{\l,g}(x,y;t)\Big],
\eeq
where the generalized nature of the propagator $G_{\l,g}$ is indicated
by the additional subscript $g$.
The graphical representation of the integral version of
eq.\ \rf{2.55} is shown in Fig.\ \ref{fig2}.
At this point, the new, generalized disc amplitude $W_{\l,g}(x)$
is unknown and has to satisfy the equation
\beq\label{3.2}
W_{\l,g} (x) = W_{\l,g} ^{(0)}(x) + 
g\int\limits_0^\infty \d t \int\limits_0^\infty \d l_1 \d l_2  \;
(l_1+l_2) G^{(0)}_{\l,g}  (x,l_1+l_2;t) W_{\l,g} (l_1)W_{\l,g} (l_2),
\eeq 
where superscripts $(0)$ indicate the CDT amplitudes introduced in
eqs. \rf{2x.1} and \rf{2.n1} above, that is,
\beq
W_{\l,g} ^{(0)}(x)\equiv W_{\l,g=0} (x)=W_\l(x),\label{relabel}
\eeq 
and similarly for $G^{(0)}_{\l,g}$. The graphical representation of eq.\ \rf{3.2}
is shown in Fig.\ \ref{fig3}.
\FIGURE[t]{
\centerline{\scalebox{0.55}{\rotatebox{0}{\includegraphics{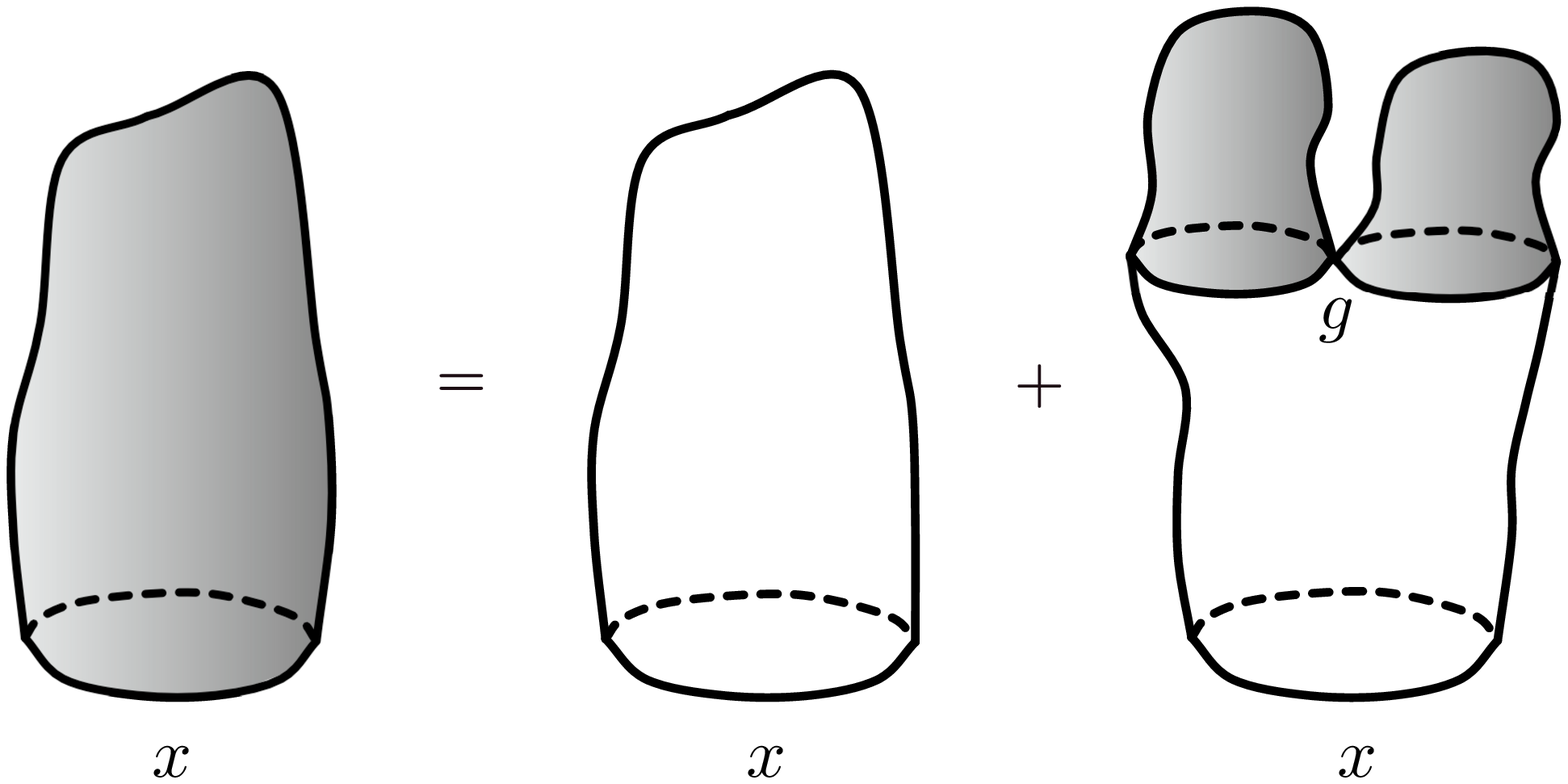}}}}
\caption[fig3]{{\small Graphical illustration of eq.\ \rf{3.2}. Shaded
parts represent the full disc amplitude $W_{\l,g}$, unshaded parts the CDT disc
amplitude $W_{\l}$ and propagator $G_\l$. 
}}
\label{fig3}
}
The integrations in \rf{3.2} can be performed and one finds \cite{alwz}
\beq\label{3.9}
\hat{W}_{\l,g}(x) = (x-c)\sqrt{(x+c)^2-\frac{2g}{c}},
~~~
c = u\sla, ~~u^3-u+\dfrac{g}{\l^{3/2}}=0,
\eeq
where 
\beq\label{3.3}
\hat{W}_{\l,g}(x)\equiv (x^2-\l) + 2g W_{\l,g}(x).
\eeq
Using the definition \rf{3.3} and eq.\ \rf{3.9}, we can write eq.\ \rf{2.55} as
\beq\label{3.5}
 \frac{\prt}{\prt t} G_{\l,g}(x,y;t) = 
- \frac{\prt}{\prt x} \Big[\hat{W}_{\l,g}(x)\, G_{\l,g}(x,y;t)\Big].
\eeq
The solution is  
\beq\label{3.6}
G_{\l,g} (x,y;t) = 
\frac{\hat{W}_{\l,g}(\bx(t,x))}{\hat{W}_{\l,g}(x)} \; \frac{1}{\bx(t,x)+y},
\eeq
where $\bx(t,x)$ is the solution of the characteristic equation for \rf{3.5}, 
\beq\label{3.7}
\frac{\d \bx}{\d t} = -\hat{W}_{\l,g}(\bx),~~~\bx(0,x)=x.
\eeq

The generalized CDT model of 2d quantum gravity we have defined above
is a perturbative deformation of the original model in the sense
that it has a convergent power expansion of the form
\beq\label{5.1}
W_{\l,g}(x) = \dfrac{1}{\sla}\sum_{n=0}^\infty c_n \Big(\frac{x}{\sla}\Big)\; 
\tg^n,~~~~\tg \equiv \left( \dfrac{g}{\l^{3/2}}\right),
\eeq 
in the dimensionless 
coupling constant $g/\l^{3/2}$. This implies in particular that 
the average number $\langle n\rangle$ of  baby universes 
created during the proper-time evolution
of the two-dimensional universe
described by this model is finite, a property already 
observed in previous 2d models with topology change \cite{lw}.
The expectation value of the number $n$ of 
branchings can be computed according to
\beq\label{5.2}
\la n\ra = \dfrac{g}{W_{\l,g}(x)}\dfrac{\d W_{\l,g}(x) }{\d g},
\eeq
which is finite as long as we are in the range of 
convergence of $W_{\l,g}(x)$.
This coincides precisely with the range where the
function $W_{\l,g}(x)$ behaves in a physically acceptable way, namely, 
$W_{\l,g}(l)$ goes to zero if the length $l$ 
of the boundary loop goes to infinity \cite{alwz}. 

Is it possible to give a gravitational interpretation of the 
new coupling constant $g$?
From a purely Euclidean point of view all graphs appearing in
Fig.\ \ref{fig3} have the fixed topology of a disc.
However, from a Lorentzian point of view, 
which comes with a notion of time, it is
clear that the branching of a baby universe is associated with a change of
the {\it spatial} topology, a
singular process in a Lorentzian space-time \cite{louko}.
One way of keeping track of this in a Wick-rotated, Euclidean picture is
as follows. Since 
each time a baby universe branches off it also has to end somewhere,  
we may think of marking the resulting ``tip" with a puncture. 
(Of course, these baby universes can in turn have baby universes branching off
them, giving rise to additional branchings and punctures.) From a
gravitational viewpoint, each new puncture corresponds to a topology
change and receives a weight $1/G_N$, where $G_N$ is Newton's constant,
because it will lead to a change by precisely 
this amount in the two-dimensional 
(Euclidean) Einstein-Hilbert action
\beq\label{5.3}
S_{EH} = -\dfrac{1}{2\pi G_N} \int \d^2\xi \sqrt{g} R.
\eeq
Let us view the  continuum theory as the limit of a lattice 
theory (CDT) with lattice spacing $a$. On the lattice we have
a dimensionless ``bare'' coupling constant
$g_0 (a) = g a^3$, where $a$ is 
the lattice spacing (see \cite{alwz} for a detailed discussion).
According to the arguments above 
we can now make the identification $g_0(a)= e^{-1/G_N(a)}$,
where $G_N(a)$ denotes the ``bare'' gravitational coupling constant. 
One can introduce a {\it renormalized}
gravitational coupling constant by
\beq\label{5.4}
\dfrac{1}{G_N^{ren}} = \dfrac{1}{G_N(a)}+\dfrac{3}{2}\ln \l a^2.
\eeq
This implies that the {\it bare} gravitational 
coupling constant $G_N(a)$ goes to
zero like $1/|\ln a^3|$
when the cut-off vanishes, $a \to 0$, 
in such a way that the product $\e^{1/G_N^{ren}}/\l^{3/2}$ is 
independent of the cut-off $a$. We can now identify 
\beq\label{5.5}
\e^{-1/G_N^{ren}} = g/\l^{3/2}
\eeq
as the genuine coupling parameter in which we expand.

This renormalization of the gravitational (or string) 
coupling constant is reminiscent 
of the famous double-scaling limit in non-critical string
theory\footnote{It is called the double-scaling limit since  
from the point of view of the discretized theory it involves a
simultaneous renormalization of the cosmological constant $\l$ and 
the gravitational coupling constant $G_N$. In this article we have already
performed the renormalization of the cosmological constant. For details
on this in the context of CDT we refer to \cite{al}.} \cite{dsl}. 
In that case one also has $g \propto e^{-1/G_N^{ren}}$, the only difference
being that relation \rf{5.4} is changed to
\beq\label{5.6}
\dfrac{1}{G_N^{ren}} = \dfrac{1}{G_N(a)}+\dfrac{5}{4}\ln \l a^2,
\eeq
whence the partition function of non-critical string theory appears  
precisely as a function  of the dimensionless coupling 
constant $g /\l^{5/4}$.

\section{String Field Theory}\label{sft}

In quantum field theory, particles can be created and annihilated 
if the process does not violate any conservation laws of the
theory. In string field theory, one has analogous operators which 
can create and annihilate strings. 
From the 2d quantum gravity point of view we are dealing with a 
third quantization of gravity: one-dimensional universes can 
be created and destroyed. In \cite{sft} such a formalism was
developed for non-critical strings in a zero-dimensional target
space (or 2d Euclidean quantum 
gravity). We will follow the procedure outlined there closely
and develop a string field theory or third quantization 
for CDT, which will allow us in principle to calculate any
amplitude involving the creation or annihilation of universes.

As starting point we assume the existence of a vacuum from
which universes can be created. We denote this state $\vac$ and
define creation and annihilation operators through
\beq\label{s1} 
[\Psi(l),\Psi^\dg(l')]=l\del(l-l'),~~~\Psi(l)\vac = \cav \Psi^\dg(l) =0. 
\eeq
This assignment corresponds to working with spatial universes where
a point has been marked. This is merely a
formal aspect, to avoid having to put in certain combinatorial 
factors by hand when gluing universes together. The operators 
$\Psi(l)$ and $\Psi^\dg(l)$ will be assigned the
dimensions ${\rm dim}\,[\Psi]={\rm dim}\,[\Psi^\dg] =0$.

We could alternatively have chosen creation and annihilation
operators which create and annihilate universes without such
a mark. Instead of \rf{s1} we then would have had 
\beq\label{s1a} 
[\Psi(l),\Psi^\dg(l')]=l^{-1}\del(l-l'),~~~\Psi(l)\vac = \cav \Psi^\dg(l) =0, 
\eeq
with corresponding dimensional assignments 
${\rm dim}\,[\Psi]=1$ and ${\rm dim}\,[\Psi^\dg] =1$. One could even
let $\Psi^\dg$ create marked universes and $\Psi$ annihilate unmarked
universes if one compensated for the missing combinatorial factors by hand.
In the following we will use the assignment \rf{s1}.

Let us write the propagator equation \rf{2.n1} using the boundary length rather 
than the
boundary cosmological constant as a variable\footnote{For convenience of 
notation we have in \rf{s2} also marked the exit loop $l_2$ in order
to have symmetry between the loops at initial
and final time, 
i.e.\ $\tG_\l(l_1,l_2;t)= l_2G_\l(l_1,l_2;t)$.}, that is,
\beq\label{s2}
\frac{\prt}{\prt t} \tG_\l(l_1,l_2;t) =  
l_1 \Big(\frac{\prt^2}{\prt l_1^2}-\l\Big) \tG_\l(l_1,l_2;t),
\eeq 
which we can also write as 
\beq\label{s3}
\tG_\l(l_1,l_2;t) = \la l_2| \e^{-t H_0(l) } |l_1\ra, ~~~~
H_0(l) =- l \frac{\prt^2}{\prt l^2}+\l l.
\eeq
Associated with the spatial universe we have a Hilbert space on the
positive half-line, and a corresponding scalar product 
\beq\label{s4}
\la \psi_1 |\psi_2\ra = \int \frac{dl}{l} \; \psi_1^* (l) \psi_2(l),
\eeq
which makes $H_0(l)$ hermitian. 
The introduction of the operators $\Psi(l)$ and $\Psi^\dg(l)$ in \rf{s1}
can be thought of as analogous to the standard second quantization 
in many-body theory. The single-particle Hamiltonian becomes in our 
case the ``single-universe'' Hamiltonian $H_0(l)$. It has normalized eigenfunctions
$\psi_n(l)$ with corresponding eigenvalues $e_n= 2n\sla$, $n=1,2,\ldots$,
\beq\label{s4a}
\psi_n(l) = l\, e^{-\sla l} p_{n-1}(l),~~~~~
H_0(l)\psi_n(l)= e_n \psi_n(l),
\eeq
where $p_{n-1}(l)$ is a polynomial of order $n\mi 1$.
We now introduce creation and
annihilation operators $a_n^\dg$ and $a_n$ corresponding to these states,
acting on the Fock vacuum $\vac$ and satisfying $[a_n,a^\dg_m]=\del_{nm}$. 
We define
\beq\label{s5}
\Psi(l) = \sum_n a_n \psi_n(l),~~~~\Psi^\dg(l) = \sum_n a_n^\dg \psi^*_n(l),
\eeq
and from the orthonormality of the eigenfunctions with respect to 
the measure $dl/l$ we recover \rf{s1}. The ``second-quantized'' Hamiltonian is
given by 
\beq\label{s6}
\hH_0 = \int_0^\infty \dll \; \Psi^\dg (l) H_0(l) \Psi(l),
\eeq
and the propagator $\tG_\l (l_1,l_2;t)$ is now obtained as
\beq\label{s7}
\tG_\l (l_1,l_2;t) = \cav \Psi(l_2) \e^{-t \hH_0} \Psi^\dg(l_1) \vac.
\eeq
\FIGURE[t]{
\centerline{\scalebox{0.70}{{\includegraphics{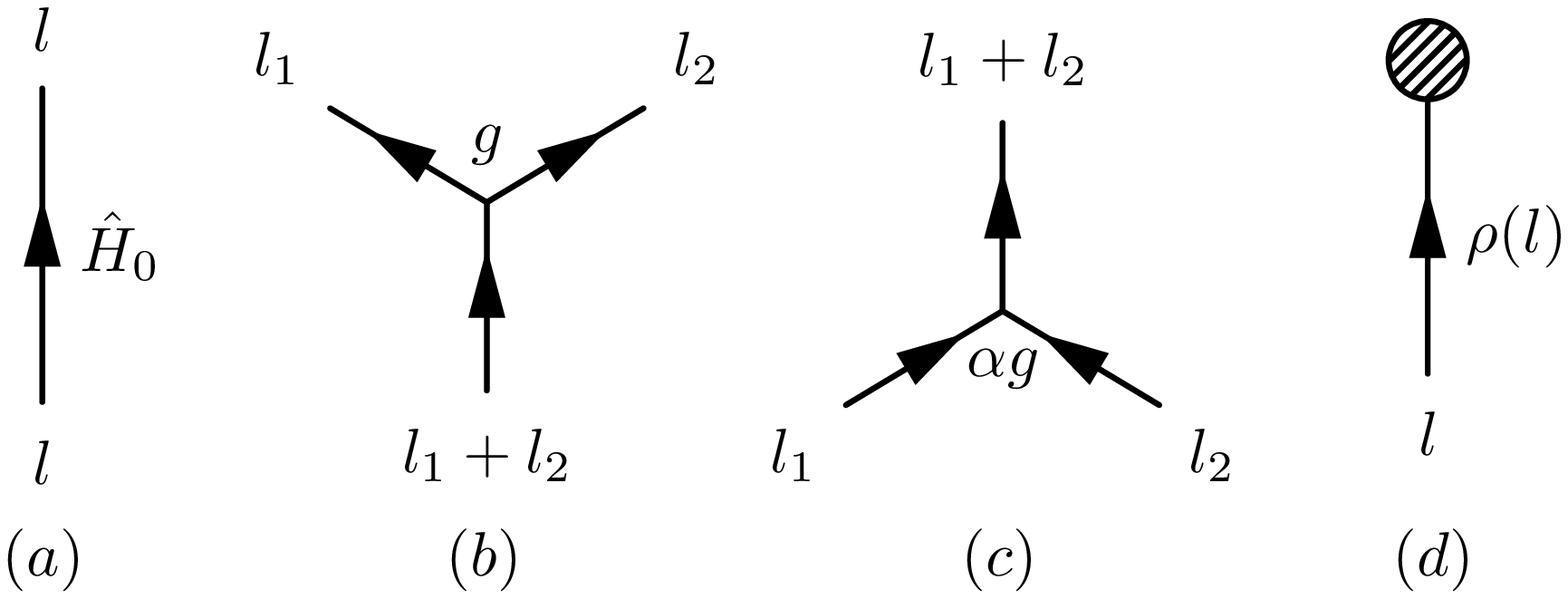}}}}
\caption[figinter]{{\small The elementary terms of the string field 
theory Hamiltonian \rf{s8}: (a) the single spatial universe propagator, 
(b) the term corresponding to splitting into two spatial universes, 
(c) the term corresponding to the merging of two spatial universes 
and (d) the tadpole term.    
}}
\label{figinter}
}
While all of this is rather straightforward, the advantage of the formalism
is that it automatically takes care of symmetry factors (like in the 
many-body applications in statistical field theory), both when many 
spatial universes are at play and when they are 
joining and splitting. Following
\cite{sft}, we define the Hamiltonian
\bea\label{s8}
\hH = \hH_0 &&-~ g \int dl_1 \int dl_2 \Psi^\dg(l_1)\Psi^\dg(l_2)\Psi(l_1+l_2)
\\ && - \a g\int dl_1 \int dl_2 \Psi^\dg(l_1+l_2)\Psi(l_2)\Psi(l_1)
-\int \dll \; \rho(l) \Psi(l), \nonumber
\eea
describing the interaction between spatial universes
(the different terms are illustrated in 
Fig.\ \ref{figinter}).
Here $g$ is the coupling constant of mass dimension 3 we have already 
encountered in Sec.\ \ref{cap}, and $\rho(l)$ denotes the amplitude for a
universe component of length $l$ to disappear into the vacuum.
The factor $\a$ has merely been introduced to distinguish between 
the action of the two terms proportional to $g$ in \rf{s8} when
expanding in powers of $g$. We will usually assume $\a=1$, unless explicitly
stated otherwise. Note that the signs of all the interaction
terms in \rf{s8} are negative. This reflects the fact that we want these 
terms to represent the insertion of new geometric structures
compared to the ``free'' propagation generated by $\hH_0$.
These structures should therefore appear with positive weight when
we expand $e^{-t \hH}$.   

The Hamiltonian $\hH$ is hermitian except for the presence of the tadpole term.
It tells us that universes can vanish, but not be created from nothing.
The meaning of the two interaction terms is as follows. The first term
replaces a universe of length $l_1+l_2$ with two
universes of length $l_1$ and $l_2$. This is precisely the 
process shown in Fig.\ \ref{fig2}. The second term represents
the opposite process where two spatial universes merge into one,
i.e.\  the time-reversed picture. The coupling constant $g$ is seen to 
appear as a string coupling constant: one factor of $g$ for 
the splitting, and another factor of $g$ for the merging of spatial
universes, leading to a combined factor $g^2$ whenever a handle
is added to the space-time.

In a way the appearance of a tadpole term is more 
natural in CDT than in the original Euclidean 
framework in \cite{sft}, since in CDT it has its origin in
a physical, causality-violating process located at the end point 
(in time) of the disc, where the baby universe disappears into nothing,
as we saw in Sec.\ \ref{cap}. The tadpole 
term is a formal realization of this. 
Because of the one-to-one correspondence between punctures and
baby universe branchings, we can also 
associate this process with the gravitational 
coupling constant, in this way linking it to $g$. 
%It was done by observing that to each splitting off of a
%baby universe we have a puncture where the baby universe ends.
%Thus the coupling constant $g$ related to the splitting of 
%spatial universes could be identified with the vanishing of 
%universes. 
The shift in associating the coupling $g$ from the splitting of
spatial universes to the vanishing of universes can be made explicit in our 
string field Hamiltonian $\hH$ in \rf{s8}.
In \rf{s8}, the coupling constant $g$ was associated with the splitting and joining
of spatial universes, but no coupling constant with 
the tadpole term, i.e.\ the vanishing of a spatial universe.
However, by redefining $\Psi$ and $\Psi^\dg$ to
\beq\label{r1}
\bar{\Psi} = \frac{1}{g} \, \Psi,~~~~\bar{\Psi}^\dg = g \, \Psi^\dg,
\eeq
the coupling constant $g$ is shifted
from the splitting to the tadpole term, i.e.\ precisely 
the shift mentioned above. In addition, the term associated
with the joining of spatial universes will have the coupling 
constant $g^2$, which precisely accounts for the change in topology.

Finally, let us identify the true, dimensionless coupling constant governing 
\rf{s8}. This can be done by re-expressing everything in units of $1/\sla$, which
represents the natural length scale of our universe. Introducing the 
dimensionless length variable $\tl = l\sla$, the dimensionless 
boundary cosmological constant $\tx = x /\sla$, the dimensionless 
time variable $\tti = t\sla$, the dimensionless tadpole density 
$\trho(\tl) = \rho(l)/\sla $, the dimensionless coupling constant 
$\tg = g/\l^{3/2}$ (already introduced in eq.\ \rf{5.1}),
and finally the dimensionless Hamiltonian $\tH = \hH/\sla$,
we can write
\beq\label{s8b}
\hH_0 = \sla \, \tH_0,~~~ 
\tH_0 = \int \frac{d\tl}{\tl}  \tPsi^\dg(\tl) H_0(\tl)\tPsi(\tl),
\eeq
where $\tPsi(\tl)=\Psi(l)$ and $\tPsi^\dg(\tl)=\Psi^\dg(l)$ 
satisfy the same commutation relation
as $\Psi(l), \Psi^\dg(l)$ when expressed in terms of $\tl$, 
and $\hH = \sla \tH$, where   
\bea\label{s8a} 
\tH = \tH_0 &&-~ 
\tg \int d\tl_1 \int d\tl_2 \tPsi^\dg(\tl_1)\tPsi^\dg(\tl_2)\tPsi(\tl_1+\tl_2)
\\ && - 
\a \tg\int d\tl_1 \int d\tl_2 \tPsi^\dg(\tl_1+\tl_2)\tPsi(\tl_2)\tPsi(\tl_1)
-\int \frac{d\tl}{\tl} \; \trho(\tl) \tPsi(\tl). \nonumber
\eea
From this expression we can read off that the true coupling constant of the 
theory is the dimensionless quantity $\tg$, precisely the ``double-scaling" 
coupling constant which already appeared in the calculation of $W_{\l,g}(x)$
and $G_{\l,g}(x,y)$, c.f. eq.\ \rf{5.1}. 
From the discussion above we also observe that
the parameter associated with a topological expansion of space-time 
is given by $\tg^2= g^2/\l^3$.
In principle we can now calculate any process which starts
from $m$ spatial universes at time 0 and ends with $n$ universes
at time $t$, represented by the amplitude
\beq\label{s100}
\tilde G_{\l,g}(l_1,..,l_m;l'_1,..,l'_n;t) =
\cav \Psi(l'_1)\ldots \Psi(l'_n) \; 
e^{-t\hH}\Psi^\dg(l_1)\ldots \Psi^\dg(l_m)\vac.
\eeq

\section{The $\mathbf{\a = 0}$ limit}\label{alpha}

\subsection{The disc amplitude}\label{disk}

Let us now consider the simplest such amplitude, that of a single 
spatial universe disappearing into the vacuum. This is 
precisely the disc amplitude of generalized CDT considered in Sec.\ \ref{cap}.
There, we allowed baby universes to branch off, but they were forbidden to
rejoin the parent universe, and thus were destined to disappear into the vacuum
eventually. In other words, the 
topology of {\it space-time} was not allowed to change during evolution. 
This can be incorporated in the string field-theoretic picture by 
choosing $\a = 0$ in \rf{s8}. The disc amplitude can then be expressed as
\beq\label{s9}
W_{\l,g}(l) =\lim_{t\to \infty}W_{\l,g}(l,t) = 
\lim_{t\to \infty} \cav \,\e^{-t \hH(\a=0)} \Psi^\dg(l)\vac.
\eeq
It describes all possible ways in which a spatial loop can develop 
in time and disappear into the vacuum without changing the topology 
of space-time. Note that the tadpole term in \rf{s8} is needed
if the amplitude \rf{s9} should be different from zero, since
the state $|l\ra = \psi^\dg (l) \vac$ is orthogonal to the vacuum state
$\vac$. We note that for $\a=0$ the vacuum expectation value
\bea\label{s9a}
\lefteqn{\cav \,\e^{-t \hH(\a=0)}\Psi^\dg(l_1)\cdots \Psi^\dg(l_m)\vac =}
~~~~~\\ 
&&\cav \,\e^{-t \hH(\a=0)}\Psi^\dg(l_1)\vac
\cav \,\e^{-t \hH(\a=0)}\Psi^\dg(l_2)\vac\cav\cdots\vac
\cav \,\e^{-t \hH(\a=0)}\Psi^\dg(l_m)\vac
\nonumber
\eea
factorizes, as one can easily prove using the algebra
of the $\Psi$'s. This is an expression of the fact that if we 
start out with $m$ spatial universes, there is no way they can merge 
at any time if $\a=0$.

Following \cite{sft}, we obtain an equation for 
$W_{\l,g}(l)$ by differentiating 
\rf{s9} with respect to $t$ and using that $\hH \vac =0$,
\beq\label{s10}
0= \lim_{t\to \infty}\frac{\prt }{\prt t}W_{\l,g}(l,t) = 
\lim_{t\to \infty} \cav e^{-t \hH(\a=0)}[\hH(\a=0), \Psi^\dg(l)]\vac.
\eeq
The commutator can readily be calculated and after a Laplace transformation
eq.\ \rf{s10} reads
\beq\label{s11}
\frac{\prt}{\prt x}\left((x^2-\l)W_{\l,g}(x)  + 
g W_{\l,g}^2(x)\right) = \rho(x),
\eeq
where the last term on the left-hand side of eq.\ \rf{s11} is a consequence 
of the factorization \rf{s9a}.
Eq.\ \rf{s11} has the generalized CDT solution \rf{3.9}-\rf{3.3}
discussed in Sec.\ \ref{cap} if 
\beq\label{s12}
\rho(x)=1,~~~{\rm i.e.}~~~\rho(l) = \del(l).
\eeq
This is a reasonable physical requirement, which we will implement 
in what follows: the spatial 
universe can only vanish into the vacuum when the length of 
the universe goes to zero.

\subsection{Inclusive amplitudes}\label{inclusive}

After reproducing the generalized CDT disc amplitude 
$W_{\l,g}(x)$ as the connected amplitude arising in the string field theory in
the limit $\a=0$, we now want to understand how
to rederive the proper-time propagator $\tG_{\l,g}(x,y,t)$ 
in this context. This propagator is characterized by an
entrance loop at time $t=0$ and an exit loop at time $t$,
and also contains baby universes which branch off and can extend
in time {\it beyond} time $t$, if only they vanish into the vacuum 
eventually, as indicated in Fig.\ \ref{fig2}.

We can reproduce this result in the $\a=0$ limit of the string field theory by introducing
the so-called ``inclusive'' Hamiltonian \cite{sft}. Since we are working 
in the $\a=0$ limit, universes can only branch and not merge
during the time evolution, and all but one have
to vanish into the vacuum. The branching process is associated with
the term
\beq\label{ds19}
g \int dl_1 \int dl_2 \Psi^\dg(l_1) \Psi^\dg(l_2) \Psi(l_1+l_2)
\eeq
in the Hamiltonian $\hH$ of eq.\ \rf{s8}. Once the branching has 
occurred, only one of the two universes can connect to the exit loop
at time $t$, the other one has to continue until it eventually
vanishes into the vacuum, a process which may occur at a
time later than $t$. This scenario is captured by replacing
\beq\label{ds20}
\Psi^\dg(l_1) \Psi^\dg(l_2) \to W_{\l,g}(l_1)\Psi^\dg(l_2)+
\Psi^\dg(l_1)W_{\l,g}(l_2)
 \eeq
in eq.\ \rf{ds19}, thus arriving at the ``inclusive Hamiltonian''
\beq\label{ds21}
\hH_{incl} = \int \dll \Psi^\dg(l) H_0(l) \Psi(l) -2g 
\int dl_1\int dl_2 \; W_{\l,g}(l_1)\Psi^\dg(l_2) \Psi(l_1+l_2),
\eeq
which enables us to rewrite the corresponding propagator $\tG_{\l,g}(l_1,l_2;t)$ as
\beq\label{ds22}
\tG_{\l,g}(l_1,l_2;t) = \cav \Psi(l_2) \,\e^{-t \hH_{incl}} \Psi^\dg(l_1)\vac.
\eeq
Differentiating eq.\ \rf{ds22} with respect to $t$, commuting $\hH_{incl}$ through to the
right and using $\hH_{incl}\vac =0$, 
one obtains after a Laplace transformation eq.\ \rf{2.55}.
We conclude that also the generalized CDT proper-time propagator has a simple 
string field-theoretic description.

\subsection{Propagator identities}\label{consistency}

Our starting point was the functional integral \rf{2.a0}
over all two-dimensional geometries with cylindrical topology, 
where the entrance and exit loop were
separated by a geodesic distance $t$. This proper-time propagator
played an important role, motivating the introduction 
of the string field Hamiltonian. As explained in footnote \ref{note1},
this construction is based on a particular definition of the geodesic distance between 
the exit and the entrance loop: every point on the exit 
loop has geodesic distance $t$ to the entrance loop, i.e.\ the 
minimal distance of a given point on the exit loop to the points 
on the entrance loop is precisely $t$, independent of the point 
on the exit loop. This implies that the exit loop as a whole has a  
specific, ``parallel" orientation relative to the entrance loop. This is a  very
useful property for the propagator to have, ensuring the existence of simple 
composition rules and thus a Hamiltonian.

What we will show next is that one can define more general amplitudes, which
depend on a somewhat looser notion of distance between their boundary
components, and which are obtained by appropriately gluing together proper-time 
propagators and disc amplitudes. These ``combined" propagators obey
non-trivial identities analogous to identities first found and verified in the 
Euclidean framework of the non-critical string field theory of \cite{sft}. 
The fact that our CDT geometries still carry some memory of their original 
Lorentzian structure after mapping them to the Euclidean sector makes 
the physical interpretation of these identities in the CDT string field theory
less clear, since the nature of the identities is rather ``Euclidean", as we shall see.

The geometric configurations we are interested in consist of two
entrance loops from which two universes propagate to the future, and then join to
form a single universe, which eventually disappears into the vacuum. Two
distinct configurations of this type are illustrated in Fig.\ \ref{fig4}. They differ
in how much time elapses in each of the ``legs" before they join. 
When summing over all geometries of fixed leg lengths $(t_1,t_2)$, the
legs will correspond to proper-time propagators of length $t_1$ and $t_2$,
and the remainder of the geometry will correspond to a disc amplitude with
boundary length $l+l'$, which has been pinched in a point such 
that it can be glued to the two exit loops of the propagators, of length $l$
and $l'$ respectively. We will be interested in comparing situations
where the two leg lengths sum to the same number $t$, such that $t_2=t-t_1$,
for different $t_1$. The left illustration in Fig.\ \ref{fig4} corresponds to the
extreme case $t_1=0$, and the right one to some intermediate choice
$t_1<t_2$. We are
not primarily concerned with the physicality or otherwise of these geometries,
but simply note that they are well defined in our string field-theoretic
set-up after Euclideanization, and possess calculable amplitudes. 
\FIGURE[t]{
\centerline{\scalebox{0.55}{\rotatebox{0}{\includegraphics{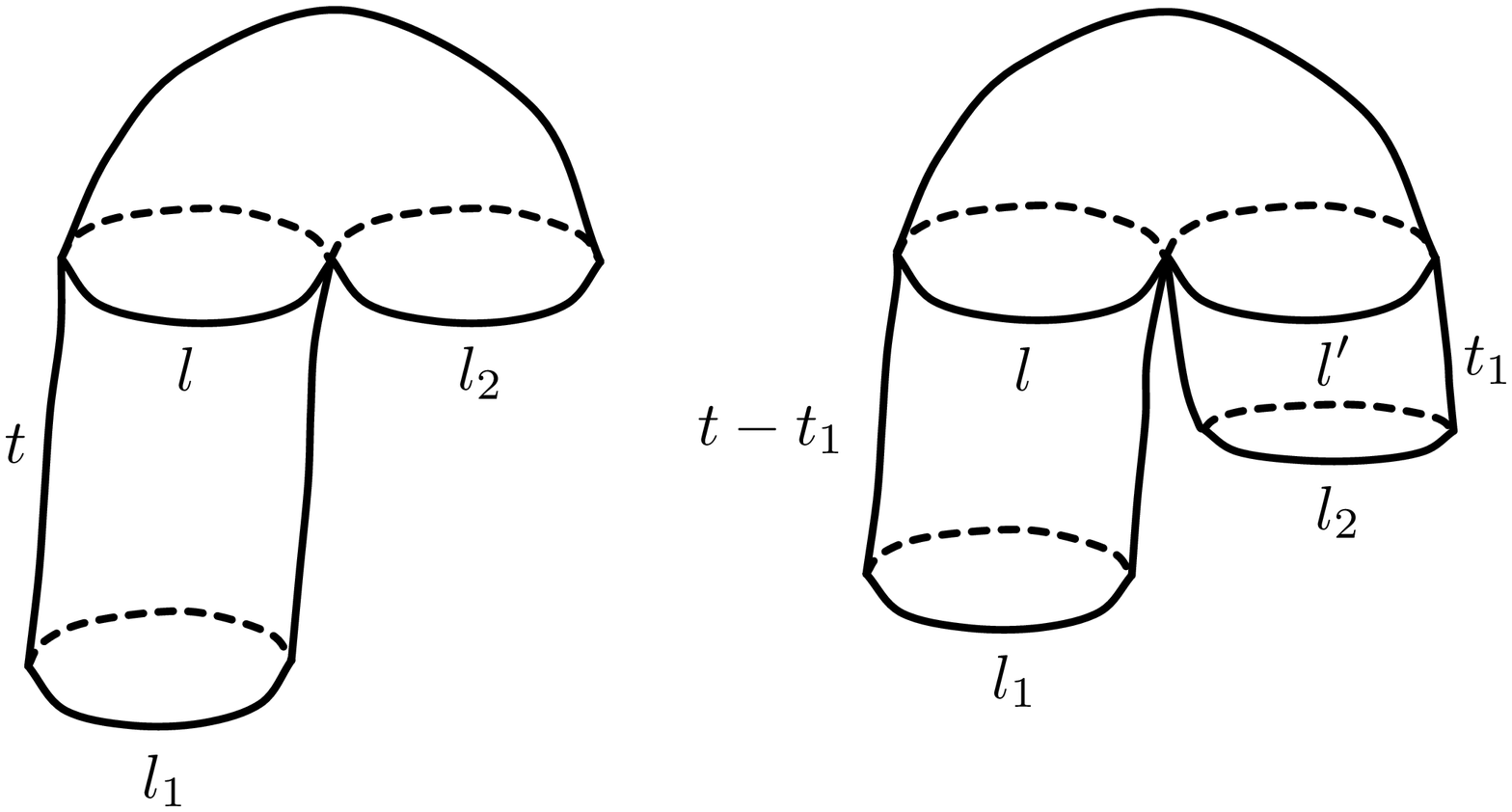}}}}
\caption[fig4]{{\small Two universes (whose time extensions add up to $t$) 
merging into a single one, and subsequently 
vanishing into the vacuum. The figure on the left shows the degenerate case 
where one leg has length $t$ and the other length 0, whereas the figure on the right
has two legs of unequal, non-zero length. An explicit computation shows that summing over
all space-times of the first type gives the same result as summing over all
space-times of the second type, for any choice of $t_1$.}}
\label{fig4}
}
We can allow the propagation to be of the most general $\a=0$ kind\footnote{Strictly
speaking, the processes described by \rf{ds23} and \rf{ds24} are of order 
$\a$ in the string field-theoretic terminology we have introduced above, since they describe
two merging spatial universes. A related amplitude $w_0(l_1,l_2)$, obtained by
integrating over all times $t$, will be introduced later, c.f. \rf{ds15c}.}, such that the dynamics is
described by the inclusive Hamiltonian $\hH_{incl}$. The two situations depicted
in Fig.\ \ref{fig4} correspond to the two calculations
\beq\label{ds23}
\int_0^\infty \d l \;  \tG_{\l,g}(l_1,l;t)  W_{\l,g}(l+l_2)
\eeq
and
\beq\label{ds24}
\int_0^\infty \d l \int_0^\infty dl'\; 
\tG_{\l,g}(l_1,l;t_1)  W_{\l,g}(l+l') \tG_{\l,g}(l_2,l';t-t_1),
\eeq
for some $0<t_1<t$. The remarkable fact is that the results of both calculations
coincide! Equivalently, one can show that
\beq\label{ds25}
0= \frac{\prt}{\prt t_1} \;\int_0^\infty dl \int_0^\infty dl'\; 
\tG_{\l,g}(l_1,l;t_1)  W_{\l,g}(l+l') \tG_{\l,g}(l_2,l';t-t_1).
\eeq
After a Laplace transformation, eq.\ \rf{ds25} reads 
\beq\label{ds26}
0=  \frac{\prt}{\prt t_1} \; 
\int_{-i\infty+c}^{i\infty +c} 
\frac{dz}{2\pi i}\; 
\tG_{\l,g}(x,-z;t_1)  W_{\l,g}(z) \tG_{\l,g}(y,-z;t-t_1).
\eeq
Using the explicit form of $G_{\l,g}(x,y;t)$, eq.\ \rf{3.6}, we 
can perform the $z$-integration in eq.\ \rf{ds26}. Next,
with the help of eq.\ \rf{3.3}, we can express $W_{\l,g}(x)$ in terms of 
$\hat{W}_{\l,g}(x)$ (given by eq.\ \rf{3.2}), and finally, using eq.\ \rf{3.7},
we can perform the $t_1$-differentiation. The result is
\beq\label{ds27}
0= \frac{\prt^2}{\prt x \prt y} \; 
\Big(\bar{x}^2(t_1,x)-\bar{y}^2(T-t_1,y)\Big),
\eeq
which is satisfied. The upshot of this calculation is that we can define a more
general amplitude 
\beq\label{ds23new}
\cG (l_1,l_2,t):=
\int_0^\infty dl \int_0^\infty dl'\; 
\tG_{\l,g}(l_1,l;t_1)  W_{\l,g}(l+l') \tG_{\l,g}(l_2,l';t-t_1)
\eeq
associated with this merger process, which
only depends on the combined distance $t$ along the legs, and which -- as we
have just proved -- is invariant under how $t$ is split into two. It is somewhat
surprising that this invariance property holds, since as Lorentzian geometries
the two situations depicted in Fig.\ \ref{fig4} are clearly distinct. Of course, during 
the Wick rotation the special character of the causality-violating
merger point between the two ``trouser legs" disappears, which may explain
the validity of \rf{ds23}, just like in the Euclidean formulation\footnote{The 
corresponding equation in the case of non-critical string theory is 
$$
0= \frac{\prt^2}{\prt x \prt y} \; \Big(\bar{x}(t_1,x)-\bar{y}(t-t_1,y)\Big),
$$
leading again to the result that the amplitude $\cG(l_1,l_2;t)$
is independent of the subdivision of $t=t_1+t_2$. Note that in this purely
Euclidean formulation a relation like \rf{ds23} appears as a (necessary) consistency 
condition, whereas in the CDT case it is satisfied as a non-trivial identity.
Moreover, we have in the Euclidean string field theory setting the additional 
consistency test that $\int_0^\infty dt \cG(l_1,l_2;t) = \cG(l_1,l_2)$, where 
$\cG(l_1,l_2)$ is the so-called universal loop-loop correlator calculated
from matrix models \cite{ajm,am}. This was verified in \cite{sft}.}. 
In line with the latter, one may interpret the quantity $\cG (l_1,l_2,t)$ as a
generalized amplitude with two boundaries separated by a distance $t$,
where the ``distance" between two spatial loops is now defined as 
the smallest geodesic distance between {\it any} pair of points 
on the two loops, with no further constraints on the relative 
position of the two loops. In particular, this makes  $\cG (l_1,l_2,t)$ symmetric under 
the exchange of $l_1$ and $l_2$.

\section{Dyson-Schwinger equations}\label{ds}

The disc amplitude $W_{\l,g}$ is one of a set of functions for which 
it is possible to derive Dyson-Schwinger equations.  
Here we will consider a more general class of functions. 
Defining the generating function $Z(J;t)$ by
\beq\label{ds1}
Z(J;t)= \cav \e^{-t \hH} \; \e^{\int dl \, J(l) \Psi^\dg(l)}\vac,
\eeq
we have 
\beq\label{ds2}
\cav \e^{-t \hH} \;\Psi^\dg(l_1)\cdots \Psi^\dg(l_n)\vac = 
\left.\frac{\del^n Z(J;t)}{\del J(l_1)\cdots \del J(l_n)}\right|_{J=0}.
\eeq
For the special case of vanishing coupling $\a =0$, we have already seen
that the amplitudes factorize, such that 
\beq\label{ds2a}
Z(J,t;\a=0) = \e^{\int dl \,J(l)W_{\l,g}(l,t)},
\eeq
where $W_{\l,g}(l,t)$ denotes the disc amplitude where the universe 
decays into the vacuum before or 
at time $t$, and where $W_{\l,g}(l,t=\infty)$ 
is the disc amplitude we have already calculated.
 
Following \cite{sft}, we can obtain the Dyson-Schwinger equations in the same 
way as for the disc amplitude, the only difference being that when 
the constant $\a$ is no longer zero, these equations do not close but
connect various amplitudes of more complicated topology. However, as we
shall see, the equations can still be solved iteratively. We denote
\beq\label{ds2b}
Z(J) \equiv \lim_{t\to \infty} Z(J;t),
\eeq
$Z(J)$ being the generating functional for universes disappearing into the
vacuum. We now have
\beq\label{ds3}
0=- \lim_{t\to \infty}\; \frac{\prt}{\prt t}\; 
\cav \e^{-t\hH} \; \e^{\int dl \, J(l) \Psi^\dg(l)}\vac =
\lim_{t\to \infty} \cav \e^{-t\hH} \; 
\hH\;\e^{\int dl \, J(l) \Psi^\dg(l)}\vac.
\eeq
Commuting the $\Psi(l)$'s in $\hH$ past the source term effectively replaces
these operators by $l J(l)$, after which they can be moved to the left of 
any $\Psi^\dg (l)$ and outside  $\cav$. 
After that the remaining $\Psi^\dg(l)$'s in $\hH$ can
be replaced by $\del/\del J(l)$ and also moved outside
$\cav$, leaving us with a integro-differential operator acting on $Z(J)$,
\beq\label{ds4}
0= \int_0^\infty dl \, J(l) \,O \left(l,J,\frac{\del}{\del J}\right)Z(J),
\eeq
where
\bea\label{ds5}
O \left(l,J,\frac{\del}{\del J}\right)&=& H_0(l) \frac{\del}{\del J(l)} -
\del (l)  \\
&& -g l \int_0^l dl'\frac{\del^2}{\del J(l')\del J(l-l')}
-\a g l\int_0^\infty dl' l'J(l')\frac{\del}{\del J(l+l')}.\nonumber
\eea
The generating functional $Z(J,t)$
also includes totally disconnected universes which 
never ``interact'' with each other. Since our main interest is 
in universes whose space-time is connected, 
the appropriate generating functional $F(J,t)$ for connected universes is
obtained by taking the logarithm of $Z(J,t)$, following standard quantum
field-theoretic methods,
\beq\label{ds6}
F(J,t) = \log Z(J,t).
\eeq
From this we obtain the correlators
\beq\label{ds7}
\cav \e^{-t\hH} \Psi^\dg(l_1)\cdots \Psi^\dg(l_n)\vac_{con} =
\left.\frac{\del^n F(J,t)}{\del J(l_1) \cdots \del J(l_n)}\right|_{J=0}, 
\eeq
and it is straightforward to 
translate the Dyson-Schwinger equation \rf{ds4}-\rf{ds5} into an equation 
for the connected functional
\beq\label{ds8}
F(J) = \lim_{t\to \infty} F(J,t),
\eeq
namely,
\bea
0= \int_0^\infty dl \, J(l)
\left\{  H_0(l)\, \frac{\del F(J)}{\del J(l)} - \delta(l) 
 -g l \int_0^l dl'\;\frac{\del^2 F(J)}{\del J(l')\del J(l-l')} \right. 
\nonumber\\
\left. -g l \int_0^l dl'
\frac{\del F(J)}{\del J(l')}\frac{\del F(J)}{\del J(l-l')}
-\a g l\int_0^\infty dl' l'J(l')\frac{\del F(J)}{\del J(l+l')}\right\}.
\label{ds9}
\eea
From eq.\ \rf{ds9} one obtains the Dyson-Schwinger equation by differentiating 
\rf{ds9} with respect to $J(l)$ a number of times and then taking $J(l)=0$.

\section{Application of the Dyson-Schwinger equation}\label{application}

Let us introduce the notation 
\beq\label{ds10}
w(l_1,\ldots,l_n) \equiv 
\left.\frac{\del^n F(J)}{\del J(l_1) \cdots \del J(l_n)}\right|_{J=0},
\eeq
as well as the Laplace transform 
\beq\label{ds11}
w(x_1,\ldots,x_n) \equiv \int_0^\infty dl_1
\cdots\int_0^\infty dl_n \; \e^{-x_1l_1-\cdots -x_nl_n} 
w(l_1,\ldots,l_n).
\eeq
Next, differentiate eq.\ \rf{ds9} with respect to $J(l)$ one, two and three
times, then take $J(l)=0$
and Laplace-transform the resulting equations. This leads to the 
following three equations (recall that $H_0(x)f(x) = \prt_x [(x^2-\l) f(x)]$):
\bea\label{ds13}
0&=&H_0(x)w(x) -1 +
g \prt_x \Big(  w(x,x) +  w(x)w(x)\Big),\\
&& ~ \nonumber\\
0&=&(H_0(x)+H_0(y))w(x,y) +g\prt_x w(x,x,y)+
g\prt_y  w(x,y,y) \label{ds15}\\
&& +2g\left(\prt_x [w(x)w(x,y)] \pl \prt_y[ w(y) w(x,y)]\right) +2\a g\prt_x\prt_y \Big(\frac{w(x)\mi w(y)}{x-y}\Big),\nonumber\\
&& ~ \nonumber\\
0&=&(H_0(x)+H_0(y) +H_0(z))w(x,y,z)\label{dsxyz} \\
&&+g\prt_x w(x,x,y,z)+
g\prt_y  w(x,y,y,z) + g\prt_z  w(x,y,z,z) \nonumber\\
&& +2g\prt_x [w(x)w(x,y,z)] + 2g \prt_y[ w(y) w(x,y,z)]+2g \prt_z[ w(z) w(x,y,z)]
\nonumber\\
&&+2g\prt_x [w(x,y)w(x,z)] + 2g \prt_y[ w(x,y) w(y,z)]+2g \prt_z[ w(x,z) w(y,z)]
\nonumber\\
&&+2\a g\left(\prt_x\prt_y \frac{w(x,z) \mi w(y,z)}{x-y} 
\pl\prt_x\prt_z \frac{w(x,y)\mi w(y,z)}{x-z}
\pl\prt_y\prt_z \frac{w(x,y) \mi w(x,z)}{y-z}\right).
\nonumber
\eea
The general structure of these equations should now be clear \footnote{Interestingly,
one can find a matrix model which reproduces the Dyson-Schwinger equations,
\cite{alwwz1}}.
We can solve the Dyson-Schwinger equations iteratively. To this end, introduce 
an expansion of $w(x_1,\ldots,x_n)$
in powers of the coupling constants $g$ and $\a$,
\beq\label{ds12}
w(x_1,\ldots,x_n) = \sum_{k=n-1}^\infty \a^k\sum_{m=k-1}^\infty g^m \; 
w(x_1,\ldots,x_n;m,k).
\eeq
The amplitude $w(x_1,\ldots,x_n)$ starts with the 
power $(\a g)^{n-1}$ since we have to perform $(n-1)$ mergings 
during the time evolution in order to create a connected geometry
if we begin with $n$ separated spatial loops. One can 
find the lowest-order contribution to $w(x_1)$ from \rf{ds13}, use that
to find the lowest-order contribution to $w(x_1,x_2)$ from \rf{ds15},
and then use this again in \rf{dsxyz}, which involves $w(x_1,x_2,x_3)$, etc. 
Returning to eq.\ \rf{ds13},
we can use the lowest-order expression for $w(x_1,x_2)$ to find the 
next-order correction to $w(x_1)$, use this and the lowest-order
correction for $w(x_1,x_2,x_3)$ to find the next-order correction
to $w(x_1,x_2)$, etc. 

Two remarks are in order: first, the integration constants arising during 
the integration of \rf{ds13}-\rf{dsxyz} and the corresponding higher-order 
equations are uniquely fixed by the requirement that the 
correlation functions fall off as the lengths $l_i \to \infty$,
i.e.\ the requirement that the Laplace-transformed amplitude 
$w(x_1,\ldots,x_n)$ be analytic for $x_i > 0$. Second, the 
expressions obtained for $w(x_1,\ldots,x_n)$ can of course be 
obtained directly from a diagrammatic expansion, using the
interaction rules shown in Fig.\ \ref{figinter}, where 
the propagation is defined by $\hH_0$, and then 
integrating in a suitable way over the times $t_i$ involved. 
The results for the first few orders are
\bea\label{g1}
w(x;0,0)& =& \frac{1}{x+\sla},\\
w(x;1,0)&=& \frac{x+3\sla}{4 \l (x+\sla)^3},
\label{g2}\\
w(x,y;1,1)&=& \frac{1}{2\sla (x+\sla)^2(y+\sla)^2},
\label{g3}\\
w(x,y,z;2,2)& =& 
\frac{7\l^{\frac{3}{2}}+5\l(x+y+z)+3\sla(xy+xz+yz) + xyz}{4\l^{\frac{3}{2}} 
(\sla+x)^3(\sla+y)^3(\sla+z)^3}.\label{g4} 
\eea
For all of these amplitudes, the space-time topology is trivial.
To lowest order in $g$,
i.e.\ without any additional baby universes, and using the results
\rf{g1}-\rf{g4} in the iteration as described above, 
the genus-one and genus-two amplitudes become
\bea\label{ds18}
w(x;2,1) &=& \frac{15\l^{\frac{3}{2}}+11\l x+
5\sla x^2+x^3}{32\l^{\frac{5}{2}}(\sla +x)^5},
\\
w(x;3,2)&=&\frac{1}{2048\l^{\frac{11}{2}}(\sla \pl x)^9}
\left(11319\l^{\frac{7}{2}} 
\pl 19951\l^3 x \pl 21555\l^{\frac{5}{2}} x^2 \pl \right.\nonumber\\ 
&&\left.  16955 \l^2 x^3 \pl 9765\l^{\frac{3}{2}} x^4 \pl 3885 \l x^5 
\pl 945 \sla x^6 \pl 105 x^7\right). \label{ds19a}
\eea
In a diagrammatic notation, the genus-two amplitude $w(x;3,2)$ corresponds 
to the following three diagrams (including suitable integrations over the times $t_i$):
\beq\nonumber
w(x;3,2)\,\,=\,\,
\raisebox{-35pt}{\includegraphics[height=80pt]{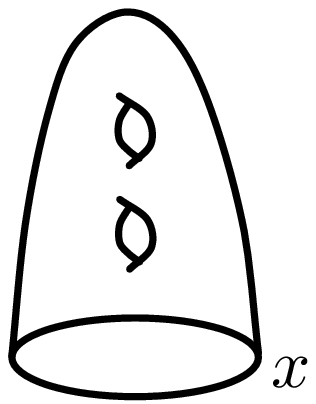}}+\,\,
\raisebox{-35pt}{\includegraphics[height=80pt]{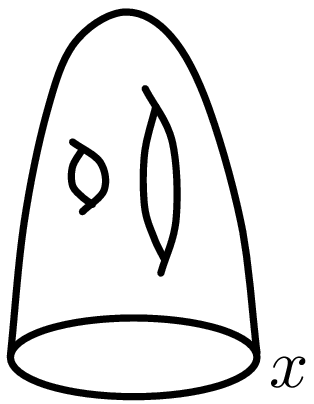}}+\,\,
\raisebox{-35pt}{\includegraphics[height=80pt]{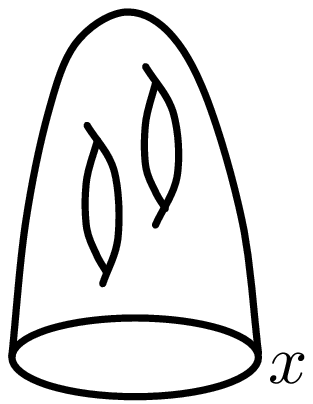}}.
\eeq

As mentioned above, the expansion of the amplitude $w(x_1,\ldots,x_n)$ starts with the 
power $(\a g)^{n-1}$, coming from merging the
$n$ disconnected spatial universes. The remaining powers
of $\a$ are associated with a non-trivial space-time topology in the form of $h$ 
additional ``handles" on the connected world sheet.
From a purely Euclidean point of view this suggest a
reorganization of the series according to
\bea\label{ds12a}
&&w(x_1,\ldots,x_n) = (\a g)^{n-1}
\sum_{h=0}^\infty (\a g^2)^h w_h(x_1,\ldots,x_n),\\
&&w_h(x_1,\ldots,x_n) = \sum_{j=0}^\infty g^j 
 w(x_1,\ldots,x_n;n-1+2h +j,n-1+h),\label{ds12b}
\eea
amounting to a topological expansion in $\a g^2$, solving at each order
for all possible baby-universe creations which at some 
point will vanish into the vacuum. 
This implies that $w_h(x_1,\ldots,x_n)$ 
will be a function of $g$, although we do not write the dependence explicitly.

The Dyson-Schwinger equations allow us to obtain the topological expansion 
iteratively, in much the same way as in our earlier power expansion in $g$. 
Since we have $w(x,x) = O(\a)$, this term does not contribute 
to lowest order; from eq.\ \rf{ds13} we 
obtain a closed equation for $w_0(x)$, namely,
\beq\label{ds13a}
H_0(x)w_0(x) + g \prt_x w_0^2(x)=1.
\eeq
This equation is of course just eq.\ \rf{s11}, where we have made the identification 
\beq\label{ds13b}
w_0(x) = W_{g,\l}(x).
\eeq
Knowing $w_0(x)$ allows us to obtain $w_0(x,y)$ from \rf{ds15},
since $w(x,y,z)$ is of order $O(\a^2)$. Therefore the three-loop term does 
not contribute to the lowest order in $\a$ of eq.\ \rf{ds15}, which is $O(\a)$,
and we find that to lowest order
\beq\label{ds15a}
\Big( H_0(x)\pl 2g \prt_x w_0(x)\pl H_0(y)\pl 2g \prt_y w_0(y)\Big)w_0(x,y)= 
\mi 2 \prt_x\prt_y \Big(
\frac{w_0(x) \mi w_0(y)}{x-y}\Big).
\eeq
We conclude that $w_0(x,y)$ is entirely determined by the knowledge of $w_0(x)$.
Note that using the definition \rf{3.3} we can simplify \rf{ds15a} to
\beq\label{ds15b}
\frac{\prt}{\prt x} \Big(\hWg(x)w_0(x,y)\Big) +
\frac{\prt}{\prt y} \Big(\hWg(y)w_0(x,y)\Big) =
- \frac{1}{g} \frac{\prt^2}{\prt x\prt y} \Big(
\frac{\hWg(x) - \hWg(y)}{x-y}\Big).
\eeq
The solution $w_0(x,y)$ can readily be found from eq.\ \rf{ds15b}, yielding
\beq\label{ds16}
w_0(x,y)= \frac{1}{f(x)f(y)}\frac{1}{4g} 
\left(\frac{[(x+c)+(y+c)]^2}{[f(x)+f(y)]^2} -1\right),
\eeq
where 
\beq\label{ds16a}
f(x) = \sqrt{(x+c)^2-2g/c} = \hWg (x)/(x-c).
\eeq
In fact, this solution was already found in \cite{alwz}
since we have by definition that
\beq\label{ds15c} 
w_0(x,y) = \int_0^\infty dt \;\cG_{\l,g}(x,y;t),
\eeq
where $\cG_{\l,g}(x,y;t)$ is the Laplace transform of 
$\cG(l_1,l_2;t)$ defined in \rf{ds24}, with $t_1=t/2$.
This function is precisely the {\it loop-loop function} of \cite{alwz}.
When expanded to lowest order in $g$, it reproduces \rf{g3}. 

As should by now be clear, one can iterate the Dyson-Schwinger equations in 
a systematic way as a power series in the number $h$ of handles
of the world sheet, exactly like we iterated them 
as a function of the coupling constant $g$, leading to
\bea\label{dse1}
w(x) &=& w_0(x) +\a g^2 w_1(x) + \a^2g^4 w_2(x)+\cdots,\\
w(x,y) &=& \a g w_0(x,y) + \a^2 g^3 w_1(x,y)+ \cdots, \nonumber
\eea 
etc. As an instructive example we will calculate the genus-one amplitude 
$w_1(x)$.
While eq.\ \rf{ds13a} was the 0th order in $\a$ of eq.\ \rf{ds13},
the 1st order reads
\beq\label{dse2}
\frac{\prt}{\prt x} \left( \hWg(x) w_1(x) + w_0(x,x) \right) =0,
\eeq
where $w_0(x,x)$ is given by eq.\ \rf{ds16}. The integration constant
is fixed by the requirement that $w_1(x)$ be analytic for
$x > 0$, i.e.\ that $w_1(l)$ fall off as $l \to \infty$. We obtain
\beq\label{dse3}
w_1(x) = \frac{w_0(c,c) -w_0(x,x)}{\hWg(x)} = 
\frac{(x+3c)(x^2+2cx+5c^2-4g/c)}{2c(4c^2-2g/c)^2((x+c)^2-2g/c)^{5/2}},
\eeq
which upon expansion in powers of $g$ to lowest order reproduces \rf{ds18},
as one would expect.

\section{Discussion}\label{discuss}

In the present work, we have developed a string field theory in zero-dimensional
target space, based on the CDT quantization 
of two-dimensional quantum gravity. It shares many properties of
the non-critical string field theory originally defined in \cite{sft}, 
from which we borrowed the formalism in the first place.
Yet, our results are different and in some ways simpler.
The tadpole term in our case is simply $\rho(l)=\del (l)$, encoding the fact 
that universes can only disappear into the vacuum if they have zero
spatial volume (that is, zero length). This is in accordance with the interaction
between spatial universes, which also preserves the total length. 
In non-critical string field theory the evolution in proper time results in
a process where the original spatial universe at proper time $t=0$
spawns an infinity of (infinitesimal) baby universes during the 
time evolution. This is related to the fact that the 
proper time in non-critical string field theory has the anomalous length dimension 
1/2. In our new CDT-based string field theory the situation is different. The proper
time $t$ has canonical dimension 1, and the number of 
baby universes created during the time evolution is finite \cite{alwz}.

It is not possible to 
connect the non-critical string field theory and the CDT-based string field theory by a 
simple analytic continuation in the coupling constant $g$,
not even in the limit as $\a=0$ \cite{alwz}. It was demonstrated in \cite{al} that,
starting from a discretized,
regularized version of the theory, the Euclidean theory 
(quantum Liouville theory) is obtained if the ``bare'' dimensionless
coupling constant $g_0$ is of order one. However, the relation between
the bare coupling constant and the dimensionful continuum coupling
constant $g$ used in the present article is given by 
\beq\label{dis1}
g_0 = g a^3,
\eeq
where $a$ is the lattice spacing in the dynamical triangulations providing 
the regularization (c.f. Sec.\ \ref{cap}). As discussed in \cite{alwz}, the generalized
CDT continuum limit corresponds to $g$ fixed, $a \to 0$, and thus 
to $g_0(a) \to 0$. The fact that $g_0(a)$ goes to zero in the
CDT string field theory is of course related to the finite number of baby universes
generated in this theory. By contrast, we have an infinite number
of baby universes generated in non-critical string field theory, where $g_0$ is 
of order one.

However, there is clearly a deeper connection between the Euclidean 
and the CDT theory awaiting to be fully understood. It was shown 
in \cite{ackl} that by integrating out the 
``excessive outgrowth" of baby universes in Euclidean 2d quantum gravity,
one recovers the CDT theory, and the mapping between the dimensionless
variables $x/\sla$ of the two theories was given explicitly. This mapping was
later discovered by Seiberg and Shih \cite{sei} as the uniformization 
map from the algebraic surface representing the ``semiclassical'' 
non-critical string to the complex plane. The singular points
of this algebraic surface correspond to so-called ZZ-branes, 
where there is a transition from compact to non-compact topology \cite{ag}. 
These singular points are mapped to points in 
the complex plane where one has a similar transition from 
compact to non-compact geometry in the CDT context \cite{awz}.
 
It would be interesting to generalize the present string field theory 
based on causal dynamical triangulations to include
the coupling to matter. In particular, one would like to investigate
whether this theory still exhibits any trace of the presence of a $c=1$
barrier. Since the existence of this barrier in the Euclidean theory 
can be partly understood as the result of an excessive creation of 
baby universes, tearing apart the two-dimensional worldsheet 
\cite{ad,adjt}, it is clear that the CDT theory may behave differently. 
Numerical simulations are compatible with the presence of a barrier
at large values of the conformal charge $c$ \cite{aal}, but no definite
results are available at this stage. Work is in progress on determining 
whether the CDT string field theory can provide a useful analytic tool
in addressing this situation. 

Equally interesting is the possibility of performing a summation 
over world sheets of all genera. Again, since the double-scaling
limit in CDT string field theory is different from the double-scaling 
limit in non-critical 
string theory, and since there is a larger ``penalty" for creating a higher-genus surface
in the sense outlined above, viewing the creation of
a higher-genus world sheet as a successive creation and annihilation
of a baby universe, one could hope that the result of such a summation 
was better behaved and less ambiguous than was the case in non-critical
string theory. Work in this direction is also in progress.

\acknowledgments{
J.A., R.L. W.W. and S.Z. acknowledge support by
ENRAGE (European Network on
Random Geometry), a Marie Curie Research Training Network in the
European Community's Sixth Framework Programme, network contract
MRTN-CT-2004-005616. R.L. acknowledges 
support by the Netherlands 
Organisation for Scientific Research (NWO) under their VICI 
program. S.Z. thanks Professor A. Sugamoto for discussion 
and the JSPS Short Term  
Award Program for financial support.}

\end{document}